\documentclass[multphys,vecphys]{svmult}

\usepackage{makeidx}     
\usepackage{graphicx}    
\usepackage{multicol}    

\makeindex             

\begin{document}

\title*{Supernovae in Galaxy Clusters}
\author{A. Gal-Yam, D. Maoz, \& K. Sharon\inst{1},
F. Prada\inst{2},
P. Guhathakurta\inst{3} \and
A. V. Filippenko\inst{4}}
\authorrunning{A. Gal-Yam et al.}
\institute{School of Physics and Astronomy,
Tel Aviv University, Israel
\texttt{avishay@wise.tau.ac.il}
\and Instituto de Astrofísica de Canarias 
and The Isaac Newton Group, Spain 
\and Herzberg Institute of Astrophysics, 
National Research Council, Canada  
\and Department of Astronomy, 
University of California at Berkeley, USA}
%
%
\maketitle

\section{Abstract}

We present the results of several surveys for supernovae (SNe) in
galaxy clusters. SNe discovered in deep, archival {\it HST} images 
were used to measure the cluster SN~Ia rate to $z=1$. A search
for SNe in nearby ($0.06 \le z \le 0.2$) Abell galaxy clusters
yielded 15 SNe, 12 of which were spectroscopically confirmed. Of
these, 7 are cluster SNe~Ia, which we will use to measure the
SN~Ia rate in nearby clusters. This search has also discovered 
the first convincing examples of intergalactic SNe. We conclude
with a brief description of ongoing and future cluster SN surveys.  

\section{Introduction}
\label{sec:1}
  
This contribution describes our observational studies of
SNe in galaxy clusters. The motivation to search 
for SNe in the fields of rich galaxy clusters is discussed 
in a contribution by Maoz et al. (this volume), and
more fully by Gal-Yam, Maoz \& Sharon \cite{GMS2002} and
Maoz \& Gal-Yam \cite{MG2003}. 
 
Observationally, SN searches in galaxy clusters were pioneered in the
late 1980's by Norgaard-Nielsen et al. \cite{NN1989}, resulting in the first
detection of a $z=0.31$ SN in the galaxy cluster AC118. More recently,
low-redshift clusters have been monitored for SNe by the Mount
Stromlo Abell Cluster SN Search \cite{Reiss1998}). 

We present below the results from several surveys for SNe in
galaxy clusters we have carried out using both ground-based
telescopes and the Hubble Space Telescope ({\it HST}).

\section{The SN rate in high-z clusters from HST}
\label{sec:2}

We have conducted a survey for high redshift SNe in deep {\it Hubble
Space Telescope} archival images of nine galaxy clusters. 
Six apparent SNe are detected (Fig. 1), with 
$21.6 \le I_{814} \le 28.4$ mag. Two SNe are associated with cluster galaxies
(at redshifts $z=0.18$ and $z=0.83$), three 
are probably in galaxies not in the clusters (at $z=0.49$, 
$z=0.60$, and $z=0.98$), and one is at unknown $z$. 
After accounting for observational efficiencies
and uncertainties (statistical and systematic) we derive the 
rate of type-Ia SNe within the projected central $500h^{-1}_{50}$ kpc of 
rich clusters: $R=0.20^{+0.84}_{-0.19} h_{50}^{2}$ SNu in 
$0.18\le z \le0.37$ clusters, and 
$R=0.41^{+1.23}_{-0.39}h_{50}^{2}$ SNu in clusters at 
$0.83\le z \le 1.27$ (95 per cent confidence interval; 
1 SNu $\equiv$ 1 SN century$^{-1}$ per $10^{10} L_{B \odot})$.
Combining the two redshift bins, the mean rate is
$R_{\bar z=0.41} = 0.30^{+0.58}_{-0.28} h_{50}^{2}$ SNu. 

\begin{figure}
\centering
\includegraphics[height=9cm]{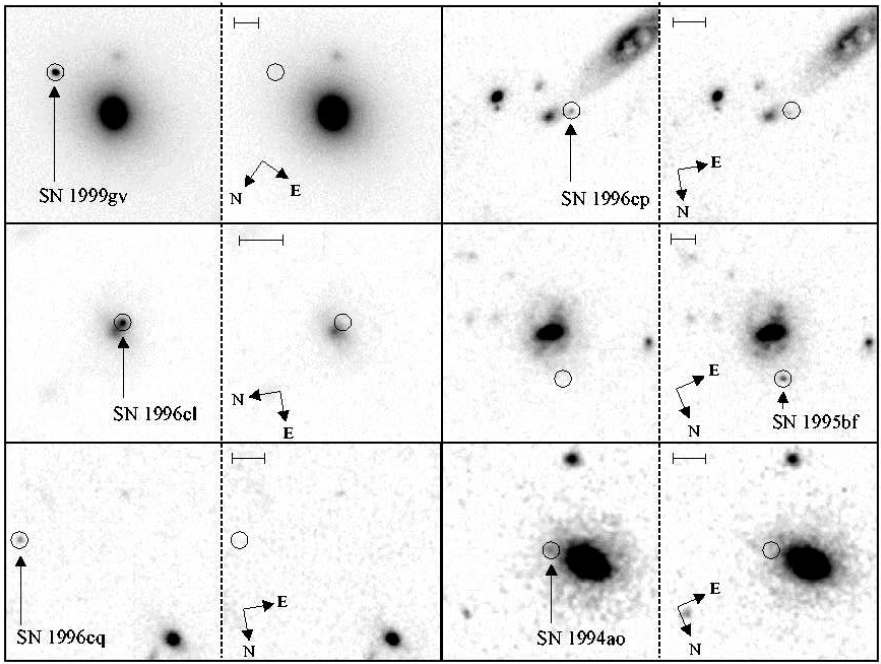}
\caption{Sections of the images, at two epochs, for each 
of the six apparent SNe discovered in our {\it HST} cluster survey. 
The scales shown in the upper-left-hand corners correspond
to $1''$.}
\label{fig:1}       
\end{figure}

We also compare our observed counts of field SNe (i.e., 
non-cluster SNe of all types) to recent model predictions.
The observed field count is $N\le 1$ SN with $I_{814}\le 26$ mag, 
and $1\le N \le 3$ SNe with $I_{814}\le 27$ mag. These counts are
about two times lower than some of the predictions. Since the 
counts at these magnitudes are likely dominated by type-II SNe, 
our observations may 
suggest obscuration of distant SNe~II, or a SN~II luminosity
distribution devoid of a large high-luminosity tail.
Further deatils are presented by Gal-Yam, Maoz \& Sharon \cite{GMS2002}. 
Additional archival SNe from {\it HST}, providing significantly
stronger constraints on the properties of high-$z$ SNe, 
are discussed in a contribution by Sharon, Gal-Yam \& Maoz (this volume).

\section{WOOTS: A survey for SNe in $0.06<z<0.2$ clusters}
\label{sec:3}

\subsection{Cluster SN rates}

The Wise Observatory Optical
Transient Search (WOOTS) is a survey for SNe
in the fields of $0.06 \le z \le 0.2$, rich Abell 
galaxy clusters using the Wise $1~$m telescope. 
15 SNe were discovered (Fig. 2), and for
12 of these we obtained follow-up spectroscopy. 
Seven SNe turned out to have
occurred in cluster galaxies, while five are field events. 
11 of the events (including all cluster SNe) are apparently SNe~Ia, 
and many were discovered near maximum light. The
cluster SN sample is suitable for the calculation of 
the SN rate in clusters (Gal-Yam \& Maoz 2003, in preparation). 

\begin{figure}
\centering
\caption{Images of the 15 supernovae discovered by WOOTS}
\label{fig:2}       
\end{figure}

\subsection{Intergalactic SNe}

Of the seven cluster SNe~Ia discovered in the course of WOOTS,
two SNe, 1998fc in Abell 403 $(z=0.10)$ and 2001al
in Abell 2122/4 $(z = 0.066)$, have no obvious hosts. 
Both events appear 
projected on the halos of the central cD galaxies, but have velocity
offsets of 750-2000 km s$^{-1}$ relative to those galaxies, suggesting they
are not bound to them. Deep Keck imaging
of the locations of the two SNe are used to 
put upper limits on the luminosities of
possible dwarf hosts, $M_R > -14$ mag for SN 1998fc and $M_R > -11.8$ mag for
SN 2001al. The fractions of the cluster luminosities in dwarf galaxies fainter
than these limits are $<3\times10^{-3}$ and $<3 \times 10^{-4}$,
respectively. Thus, $2/7$ of the SNe would be associated with
$\le3\times10^{-3}$ of the luminosity attributed to galaxies. 
It is argued, instead, that the progenitors of both
events were probably members of a diffuse population of intergalactic stars,
recently detected in local clusters via planetary nebulae and red giants.
Considering the higher detectability of hostless SNe compared to normal SNe, we
estimate that $20^{+12}_{-15}$ percent of the SN~Ia parent stellar population
in clusters is intergalactic. This fraction is consistent with other
measurements of the intergalactic stellar population, and implies that the
process that produces intergalactic stars (e.g., tidal disruption of cluster
dwarfs) does not disrupt or enhance significantly the SN~Ia formation
mechanism. Hostless SNe are potentially powerful tracers of the formation of
the intergalactic stellar population out to high redshift. Further details
can be found in Gal-Yam, Maoz, Guhathakurta, \& Filippenko \cite{GMGF2003}.  

\section{Future Prospects}
\label{sec:4}

Measurements of SN rates in clusters out
to $z=1$ provide a powerful and unique tool to probe cluster
metal enrichment and the progenitors of SNe~Ia (see
contribution by Maoz et al., this volume). 
The results are currently limited by large
errors due to small number statistics. It is therefore
desirable to obtain additional measurements of SN rates
in clusters. The WOOTS cluster SN sample will provide another
measured point at $z\sim0.15$. 
We have also begun a ground-based survey for SNe in rich, lensing
clusters at $z\sim0.3$, using the $2.5~$m NOT telescope
at La Palma, Spain. This survey aims to enlarge the number
of SNe used to calculate the cluster SN rate in the $z\sim0.3$ bin
(1 SN) by an order of magnitude, and thus to significantly
decrease the statistical error. The first results of this 
survey include the discovery of several cluster SNe 
(e.g., Fig. 3; Gal-Yam, Maoz, Prada \& Guhathakurta 2003\cite{gmpg2003})
as well as new strong lensing clusters.
A similar survey for SNe in high      
redshift clusters ($z\sim0.8$) using $4~$m class (or larger) 
telescopes is being planned. With such data, 
it will be possible to set strong constraints
on the origin of iron in the ICM and the characteristic delay time
of SNe~Ia.

\begin{figure}
\centering
\includegraphics[height=7cm]{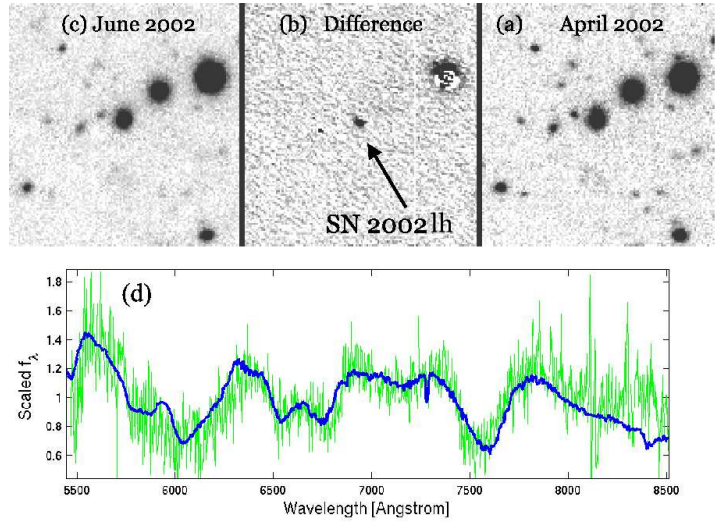}
\caption{Subtraction of a NOT image of Abell 1961 (z=0.232)
obtained in April 2002 
(a) from a similar image obtained in June 2002 (c) reveals SN 2002lh (b). In panel
(d) the comparison of the Keck spectrum of this event (light) with a redshifted
spectrum of the nearby SN 1999ee near peak magnitude (bold, from Hamuy et al. \cite{H2002}),
reveals this is a SN~Ia at z=0.236, i.e., in one of the cluster galaxies.}
\label{fig:3}       
\end{figure}

\printindex
\end{document}